\documentstyle[12pt,epsf]{article}

\catcode`\@=11
\newbox\tempboxa
\newdimen\captionboxsubcount
\def\capsize#1{\captionboxsubcount=#1pt}
\newdimen\captionboxsub
\captionboxsub=\hsize \advance\captionboxsub by -\captionboxsubcount
\advance\captionboxsub by -\captionboxsubcount
\long\def\@makecaption#1#2{
 \setbox\@tempboxa\hbox{#1: #2}
 \ifdim \wd\@tempboxa >\captionboxsub
\rightskip=\captionboxsubcount \leftskip=\captionboxsubcount #1: #2
\else \hbox to\hsize{\hfil\box\@tempboxa\hfil}
 \fi}
\catcode`\@=12
\capsize{30}

\begin{document}

\begin{titlepage}
\begin{flushright}
OU-HET 283 \\
hep-th/9711177 \\
November 1997
\end{flushright}
\bigskip
\bigskip
\bigskip
\begin{center}
{\Large \bf 
GEOMETRICAL ANALYSIS OF \\
BRANE CREATION VIA $M$-THEORY
}
\bigskip
\bigskip
\bigskip
\bigskip
\bigskip

{\Large Yuhsuke Yoshida}
\bigskip

{\it
Department of Physics,\\
Graduate School of Science, Osaka University,\\
Toyonaka, Osaka 560, JAPAN
}
\end{center}
\bigskip
\bigskip
\bigskip
\begin{abstract}
{\normalsize
A geometrical analysis is given of Dirichlet fourbrane creation, when
sixbrane crosses fivebrane in $M$-theory.
A special property of the Taub-NUT space leads to the consequence.
When brane configurations are considered for four dimensional
${\cal N}=2$ field theories, sixbrane contributes to the
$\beta$-function through the Dirac string of the Taub-NUT space.
}
\end{abstract}
\end{titlepage}

\section{Introduction}

Brane creation is a key to understanding string dualities and
also field theory dualities.
After the first analysis\cite{HW}, many discussions and
results\cite{Anomaly,D0D8,D0O6,D4D4,NOYY2} are given for brane
creations.
In this paper, we show a proof of brane creation\cite{NOYY2} and give
a new result relating to Dirichlet sixbrane.
An importance is stressed of the role of the Dirac string from
Dirichlet sixbrane.

In section~\ref{BC} we refine our proof in Ref.~\cite{NOYY2}.
We consider, in $M$-theory, the simplest case of an NS fivebrane and a 
Dirichlet sixbrane, since it is enough to understand the Dirichlet
fourbrane creation when the two branes cross.
We figure out shapes of the NS fivebrane changing the
distance to the Dirichlet sixbrane.
We find a crossover from a smooth to a distorted NS fivebrane.
Then, we take the type IIA limit of the $M$-theory brane
configurations, and we show that a Dirichlet fourbrane is created
under a suitable situation.
In section~\ref{FDS} we consider an $M$-brane configuration for four
dimensional ${\cal N}=2$ supersymmetric QCD.
We investigate how the simplest analysis, given in section~\ref{BC},
works in this brane configuration.
We find a difficulty that for a specific crossing of branes an unknown 
fourbrane is created.
To save the difficulty the Dirac string extending from the Dirichlet
sixbrane plays an important role.
In order to see that the Dirac string is not a unphysical object at
all, we show that the Dirichlet sixbrane contributes to the
$\beta$-function of the low energy field theory through the Dirac
string.

\section{Brane Creation}\label{BC}

We consider an $M$-theory brane configuration according to
Ref.~\cite{Witten}.
We are assuming that the circumference of the $11^{th}$
dimensional space is $4\pi R$.
We put $R=1$ for simplicity, otherwise stated.
We start by considering a single fivebrane and a single sixbrane.
Sixbrane is meant by a removable NUT singularity of a Taub-NUT space
$Q$ in $M$-theory.\cite{Townsend}
Then, putting the sixbrane in the point $(v,b) = (0,b_0)$ defines the
Taub-NUT space $Q$.
The worldvolume of the fivebrane is topologically ${\bf R}^4\times
{\bf C}$.
The four dimensional part ${\bf R}^4$ is parametrized by the
coordinate $(x^0, x^1, x^2, x^3)$.
The complex plane ${\bf C}$ is embedded holomorphically into the
Taub-NUT space, by which the ${\cal N}=2$ supersymmetry is left in
four dimensions.
We introduce complex coordinates
\begin{equation}
v =
\frac{x^4+ix^5}{R} ~, \quad
s = b + i\sigma =
\frac{x^6+ix^{10}}{R} ~.
\end{equation}
One choice of holomorphic coordinates\cite{NOYY} of the Taub-NUT space
are $v$ and
\begin{equation}\label{y}
y = e^{-s/2}
\left( -b+b_0+
   \sqrt{(b-b_0)^2+|v|^2} \right)^{\frac{1}{2}} ~.
\end{equation}
If the $U(1)$ gauge symmetry is fixed in such a way that the Dirac
string extends from the sixbrane to the positive $x^6$ direction,
we have the above holomorphic coordinate (\ref{y}).
Here, Dirac string severely restricts the way to take a holomorphic
coordinate, and vice versa.
If we define a part of the holomorphic coordinates by setting
$v=x^4+ix^5$, then there are only two ways to put the Dirac string in
$(x^4,x^5,x^6)$ space; one way is to put it so that it extends from
sixbrane to the positive infinity of the $x^6$ direction, and the
other way is to the negative infinity.

\subsection{Geometry of single fivebrane and sixbrane}

Now, we come to the point.
According to Witten's argument\cite{Witten}, there are two distinct
ways of embedding in the present setup, depending on whether the
sixbrane is to the left or the right of the fivebrane.
We will consider the case when the sixbrane is to the left of the
fivebrane.
Thus, we pick up the holomorphic coordinate $y$ defined by
eq.~(\ref{y}) and embed the complex plane ${\bf C}$ by
\begin{equation}\label{M5}
y = v ~.
\end{equation}

The angular part of eq.~(\ref{M5}) states that $\sigma$ jumps by
$\pm 4\pi$ when one circles around the origin in the complex $v$
plane.
This requires an interpretation:
The Taub-NUT space $Q$ is a circle bundle on the three dimensional
base space ${\bf R}^3\setminus0$, where $0$ is the position of the NUT 
singularity.
We have a projection $\pi$ to the base space;
$\pi:Q\to{\bf R}^3\setminus0$.
The projection $\pi$ is defined by forgetting the dependence of the
variable $\sigma$.
Suppose we project, by $\pi$, the curve embedded in this fibre bundle
into the base space.
The radial part of the embedding equation represents the projected
curve, and the projected curve is again two dimensional.
Define a closed one-cycle on the projected curve on the base space.
The value of $\sigma$ can jump by $\pm4\pi$ when we go along the
one-cycle, which is meant by the fivebrane is ``wrapping'' around
$x^{10}$.
We should note that in a usual type IIA language such ``wrapping''
means by a non-trivial $U(1)$ gauge potential on the NS fivebrane
worldvolume.
Let us see that the jump $\sigma\to\sigma\pm4\pi$ implies a vortex
on the fivebrane worldvolume.
We have a $U(1)$ gauge potential $w_i$ and a scalar $\sigma$ on the
two dimensional part ${\bf C}$ of the fivebrane worldvolume.
These two fields are related by the $U(1)$ gauge symmetry;
$\delta w_i = -\partial_i a$ and $\delta\sigma = a$.
This symmetry defines the minimal coupling between $w_i$ and $\sigma$.
In addition, we have anomalous coupling
$\sigma\epsilon^{ij}\partial_i\omega_j$ with an identification
$\sigma\sim\sigma+4\pi$.
Now, the NS fivebrane is penetrated by the Dirac string of the
sixbrane in the present coordinate system.
Then, picking up a small two dimensional disc $\delta D$ from the two
dimensional part ${\bf C}$ of the fivebrane around the origin
$v\sim0$, we obtain 
\begin{equation}
\int_{\delta D} d\omega = 4\pi ~,
\end{equation}
which induces a vortex on the NS fivebrane worldvolume.

The radial part of eq.~(\ref{M5}) determines the form of the complex
plane ${\bf C}$ in the base space of the Taub-NUT space as
\begin{equation}\label{vb}
|v| = e^{-b/2} \sqrt{e^{-b}-2(b-b_0)} ~.
\end{equation}
Since this equation has a $U(1)$ symmetry rotating $v$,
we may regard the variable $v$ as non-negative real number
($v\in{\bf R}_{\ge0}$) without loss of generality.
The maximum value of $b$, which we call $b_1$, is given at the
intersection point of the fivebrane and the Dirac string extending
from the sixbrane to the right-infinity.
Then, the value of $b_1$ is determined by
\begin{equation}\label{b01}
b_0 = b_1 - \frac{1}{2}\exp(-b_1) ~,
\end{equation}
which is the zeros of eq.~(\ref{vb}).
With the relation (\ref{b01}) at hand, instead of considering the
quantity $b_1$ as a function of $b_0$, it is often convenient to
regard $b_0$ as a function of $b_1$.
As explained in the following, the relation allows us to capture
briefly a first observation that when the sixbrane crosses the
fivebrane a ``new'' fivebrane is created.
A complete proof will appear in subsection~\ref{IIA}.
If $b_0\gg1$, the first term in eq.~(\ref{b01}) dominates, and we
have $b_1 \cong b_0$, which shows that the top of the fivebrane
responds linearly to the position of the sixbrane.
If $b_0\ll-1$, the second term in eq.~(\ref{b01}) dominates, and we
have $b_1 \cong -\ln(-2b_0)$, which means that $b_1$ does not
depend on $b_0$ so much.
The intermediate region $b_0\sim0$ is a crossover between these two
behaviors.

Now, let us draw the geometrical shape of the embedded fivebrane from
eq.~(\ref{vb}) in more detail.
As learned in the above briefly, clear situations appear when the
sixbrane is far from the origin.
First, let us consider the case $b_0\gg1$.
The functional form of $v$ asymptotically becomes
\begin{equation}\label{b<b0}
v = e^{-b/2}\sqrt{2(b_0-b)} ~,
\end{equation}
for $b<b_0$.
This shows that $b$ approaches to $b_0$ if $v$ goes to zero.
Next, let us consider the case $b_0\ll-1$.
Asymptotically we have
\begin{equation}\label{b1<b}
v = e^{-b_1/2}\sqrt{2(b_1-b_0+1)}\cdot\sqrt{b_1-b} ~,
\end{equation}
for $b<b_1$.
The fact that $b_1 \ll -1$ implies the numerical factor in front of
eq.~(\ref{b1<b}) is large, and then a small variation of $b$ around
$b_1$ leads to a large variation of $v$.
In other words, the quantity $b$ is almost independent for the value
of $v$.
Namely, we have asymptotically $b=b_1$ for ${}^\forall v$.
Notice that the asymptotic form (\ref{b1<b}) with $b_0<b$ is also
valid for the former case $b_0\gg1$.

As a result, the following situations are realized geometrically:
When $b_0\to+\infty$, a part of the fivebrane gets involved in the
distant sixbrane, and is wrapping around the $11^{th}$ dimension.
This involved part of the fivebrane may be interpreted as a
semi-infinite Dirichlet fourbrane of long thin tube extending to the
right.
On the other hand, when $b_0\to-\infty$, the fivebrane becomes a
Neveu-Schwartz fivebrane parallel to the $v$ plane.
In these two limits, the Taub-NUT space becomes a flat space $Q_0 =
{\bf R}^3\times{\bf S}^1$ around the origin.
This result will give a geometrical explanation of a brane creation
via $M$-theory.
However, we have no clear distinction of the boundary between NS
fivebrane and D fourbrane.
The best way to figure out such distinction is to take the type IIA
limit in the $M$-theory, which will be done in the next.

\subsection{The type IIA limit}\label{IIA}

In this section we recover the parameter $R$ and we put $x=x^6$ and
redefine $v=x^4+ix^5$ for simplicity.
The IIA limit is defined as $R\to0$ for a fixed position of the
sixbrane $x_0 = b_0R$.
As the parameter $R$ approaches to zero, the shape of the fivebrane
changes, and we have essentially two final shapes depending on the
value of the sixbrane position $x_0$.
The following analysis will draw the final two shapes.

First of all, let us consider how the bended tail, i.e. around
$|v|\sim\infty$, of the fivebrane becomes.
In this region we have $b\sim-\infty$, and from eq.~(\ref{vb}) we
obtain asymptotically $v = R e^{-x/R}$.
Taking inverse in this asymptotic relation and taking the IIA limit,
we find $x=0$ for any large value of $|v|$.
This means that the tail of the fivebrane is fixed to be $x=0$.

Second, we consider the relation (\ref{b01}).
The crossover point $b_1=b_c$ is evaluated from $2b_c=\exp(-b_c)$, and 
we have $0<b_c<1/2$ or numerically $b_c=0.352\cdots$.
The important point is that this value is a finite constant.
Then, in the IIA limit, we have the crossover point
$x_c\equiv\lim_{R\to0} b_cR = 0$.
Moreover, a slight consideration about eq.~(\ref{b01}) allows us to
obtain
\begin{equation}\label{x01}
x_1 = (x_0-x_c)\; \theta(x_0-x_c) ~,
\end{equation}
in the IIA limit.
This relation clearly shows that if $x_0<x_c$ the position $x_1$ of
the fivebrane stays and if $x_c<x_0$ the position $x_1$ moves with
$x_0$.

Finally, we figure out from eq.~(\ref{vb}) how the shape of the
fivebrane becomes geometrically in the IIA limit.
Depending on the values of $x_0$ there are three ways to take the IIA
limit; such as $b_0\to+\infty$, $b_0=0$ and $b_0\to-\infty$ for
$x_0>0$, $x_0=0$ and $x_0<0$, respectively.
Then, to take the IIA limit of the functional form (\ref{vb}), we
divide the region of $x_0$ into the three pieces;
(a) $x_0>0$, (b) $x_0=0$ and (c) $x_0<0$.

\begin{figure}[htbp]
\epsfxsize=16cm
\begin{center}
\ \epsfbox{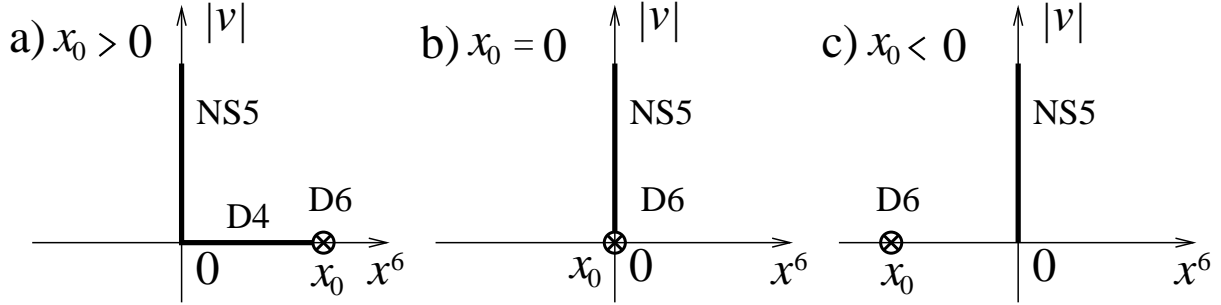}
\caption[]{
There are three brane configurations in the type IIA limit, depending
on the value of $x_0$.
In the case (a) a Dirichlet fourbrane is created suspending between
the NS fivebrane and Dirichlet sixbrane.
Here we put $v=x^4+ix^5$.
}
\label{IIA branes}
\end{center}
\end{figure}

\subsection*{(a) $x_0>0$}
In this case we send $b_0\to+\infty$ in the IIA limit, and
as well as we have $b_1=b_0\to+\infty$ asymptotically.
Then, we obtain $x_1 = \lim_{R\to0}b_1R = x_0$.
We easily find that brane in the region $x_0\le x\le x_1$ shrinks to
the point $(v,x)=(0,x_0)$ in the IIA limit.
It is enough to consider the region $x\le x_0$ for the present case.
Let us find the functional form between $x$ and $v$ in the IIA limit
in a region wise manner, dividing the region of $x$ into three;
$x=x_0$, $0\le x<x_0$ and $x\le0$.
If $x=x_0$, we have $v=0$ from eq.~(\ref{vb}) in the IIA limit.
If $0\le x<x_0$, we also have $v=0$ from eq.~(\ref{vb}), or
eq.~(\ref{b<b0}), in the IIA limit.
Therefore, as approaching $R\sim0$, the fivebrane in $x>0$ becomes a
thin tube, and finally a straight line in the IIA limit reducing its
dimension by one.
If $x\le 0$, the functional form (\ref{vb}) approaches to
\begin{equation}
v = R e^{-x/R} ~.
\end{equation}
Taking the inverse of this form, we find
\begin{equation}
x = 0 ~,
\end{equation}
for ${}^\forall v \ne 0$ in the IIA limit.
The resultant brane configuration is depicted in
Fig.~\ref{IIA branes} (a).
How about the variable $\sigma$?
The angular part of eq.~(\ref{M5}) states in the IIA limit 
that there is a vortex on the resultant NS fivebrane at the origin of
the $v$ plane.
It may be interpreted that this vortex is made by the fourbrane ending 
on the NS fivebrane.
Moreover, as explained before, the jump of $\sigma$ by $\pm4\pi$ means 
that the thin tube, or the straight line $[0,x_0]$,
in ${\bf R}^3\setminus0$ is found wrapping around the $11^{th}$
dimension in Q.
Now, we prove that a Dirichlet fourbrane is created when a sixbrane
crosses a NS fivebrane.

\subsection*{(b) $x_0=0$}
In this case we have $x_1 = \lim_{R\to0}b_cR = 0$.
So, we are enough to consider in the region $x\le0$.
For the parameter $R$ goes to zero, the functional form (\ref{vb})
asymptotically approaches to $v=\exp(-x/R)/R$.
Taking the inverse to this form, and the IIA limit, then we find
$x=0$ for ${}^\forall v$.
The resultant brane configuration is depicted in
Fig.~\ref{IIA branes} (b).

\subsection*{(c) $x_0<0$}
In this case we send $b_0\to-\infty$ in the IIA limit.
{}From eq.~(\ref{b01}) we have
$x_1 = \lim_{R\to\infty}Rb_1 = - \lim_{R\to\infty}R\ln(-2x_0/R) = 0$.
The functional form (\ref{vb}) asymptotically approaches to
\begin{equation}
v = e^{-b_1/2} \sqrt{2(x_1-x_0+R)}\cdot\sqrt{x_1-x} ~.
\end{equation}
After $x$ is expressed in terms of $v$, we find, in the IIA limit,
\begin{equation}
x = x_1 - \frac{e^{b_1}}{2(x_1-x_0+R)}v^2 \to 0 ~,
\end{equation}
for ${}^\forall v$.
The resultant brane configuration is depicted in
Fig.~\ref{IIA branes} (c).
In this case, as well as the case (b), the NS fivebrane has a vortex
at the origin of the $v$ plane, however, there is no D fourbrane.
Then, it is impossible to understand that the vortex is stemmed from
any D fourbrane.
Instead, we should actually interpret this vortex as a consequence of
the fact that the fivebrane is penetrated by the Dirac string coming
from the sixbrane.
This point will be explained in the next section.

\section{Fivebranes With Dirichlet Sixbrane}\label{FDS}

In this section we consider the brane configuration for four
dimensional finite ${\cal N}=2$ $SU(n_c)$ supersymmetric QCD.
The general setup of Witten\cite{Witten} is used here.
We will observe a peculiar property stemming from Dirac string of the
multi-Taub-NUT space.
We put $n_c$ Dirichlet fourbranes on $v = \phi_a$ $(a=1,\cdots,n_c)$
and $2n_c$ Dirichlet sixbranes on $(v,b) = (e_j,b_j)$
$(j=1,\cdots,2n_c)$, where we do not consider the center of mass;
$e_{_S}=0$.
Let us consider a generic brane configuration in which any fourbranes
and sixbranes do not collide, which means $e_i\ne e_j$ ($i\ne j$) and
$\phi_a \ne e_j$.
Setting a value of the SQCD coupling constant defines the following
parameter $h$, and the Riemann surface for the $M$-fivebrane is
\begin{equation}
y^2 - 2B(v)y+C(v) = 0 ~,\label{SWc}
\end{equation}
where
\begin{eqnarray}
B(v) &=& \prod_{a=1}^{n_c}(v-\phi_a) ~,\quad \sum_a \phi_a = 0 ~, \\
C(v) &=& -h(h+2)\prod_{j=1}^{2n_c}(v-e_j-h\cdot e_{_S}) ~.
\end{eqnarray}
The sixbranes define the multi-Taub-NUT space with the metric
\begin{eqnarray}
ds^2 &=& \frac{V}{4}|d\vec{r}|^2 +
         \frac{V^{-1}}{4}(d\sigma+\omega)^2 ~, \quad
\vec{r} = (x^4,x^5,x^6) ~, \quad \sigma = x^{10} ~, \\
V &=& 1 + \sum_{j=1}^{2n_c} \frac{1}{|\vec{r}-\vec{r_j}|} ~.
\end{eqnarray}
The one-form $\omega=\omega_idr^i$ satisfies the Bogomolny equation
for the Dirac monopoles;
$\epsilon_{ijk}\partial_j\omega_k = \partial_i V$.

The multi-Taub-NUT space is parameterized by two coordinates $v$ and
$y$.
As mentioned in section~\ref{BC}, choosing one holomorphic coordinate
$v$, say $v=x^4+ix^5$ as above, severely restricts the direction of
the Dirac strings coming from the sixbranes.
With this choice of $v$ only the positive or negative direction of the 
coordinate $b=x^6$ is allowed for the Dirac strings.
If the Dirac string extends to the ``right'', the one-form is
\begin{equation} \label{w+}
\omega = \omega_+ \equiv \sum_j\frac{(r^2-r^2_j)dr^3-(r^3-r^3_j)dr^2}
{|\vec{r}-\vec{r_j}|(-r^1+r^1_j + |\vec{r}-\vec{r_j}|)} ~,
\end{equation}
and the other holomorphic
coordinate is
\begin{equation}\label{yp}
y = e^{-s/2} \prod_{j=1}^{2n_c}
               \sqrt{-b+b_j+\sqrt{(b-b_j)^2+|v-e_j|^2}} ~,
\end{equation}
with $s = b+i\sigma=x^6+ix^{10}$.
If the Dirac string extends to the ``left'', the one-form is given by
eq.~(\ref{w+}) with $\vec{r}-\vec{r_j}$ changed the sign, here we call
it $\omega = \omega_-$, and the other holomorphic coordinate is
\begin{equation}\label{ym}
y = e^{+s/2} \prod_{j=1}^{2n_c}
               \sqrt{b-b_j+\sqrt{(b-b_j)^2+|v-e_j|^2}} ~.
\end{equation}

\subsection*{More on brane creation}

Since we would like to know the behavior of branes when a sixbrane
approaches to the fivebrane along the $b$ direction,
it is enough to concentrate on the single sixbrane, say $v\sim e_1$.
We obtain two branches $y_\pm = B \pm \sqrt{B^2-C}$ from the curve
(\ref{SWc}).
In the leading order of $v-e_1$ we find $y_+=\alpha$ and
$y_-=\beta(v-e_1)$ for some non-zero constants $\alpha$ and $\beta$.
Picking up one, $(v,y)$, of the two possible holomorphic coordinate
systems, the two NS fivebranes are embedded by $y=y_+$ and $y=y_-$.
First of all, let us consider the case when the Dirac string extends
to the right; i.e. $w=w_+$.
In the leading order of $v-e_1$ the holomorphic coordinate becomes
\begin{equation}
y = \lambda e^{-s/2} \sqrt{-b+b_1+\sqrt{(b-b_1)^2+|v-e_1|^2}} ~,
\end{equation}
for some non-zero constant $\lambda$.

Which branches $y_\pm$ is the right NS fivebrane corresponding to?
A little effort reveals that the branch $y_-$ is embedded to the right
of the branch $y_+$ in this case.
Then, the branch $y_-$ is corresponding to the right NS fivebrane.
The analysis of the embedding $y=y_-$ is exactly the same as that
given in section~\ref{BC}.
As for the embedding $y=y_+$, the analysis is carried out almost the
same way as for the case $y=y_-$.
Taking the radial part of $y=y_+$ projects the curve into the base
space ${\bf R}^3\setminus0$.
The radial part of $y=y_+$, rewritten as
\begin{equation}
|v-e_1| = \left|\frac{\alpha}{\lambda}\right| e^{b/2}
\sqrt{\left|\frac{\alpha}{\lambda}\right|^2e^b+2(b-b_1)} ~,
\end{equation}
states the same result of the case $y=y_-$:
When the sixbrane approaches to the branch $y_+$, the fivebrane is
fairly distorted.
When the sixbrane goes away from the branch $y_+$, the fivebrane
remains its shape.

Next, let us consider the angular part of $y=y_+$.
The angular part states that the $\sigma$ does not jump when one
circles around the position $v=e_1$, or rather the $\sigma$ is a
constant.
Namely, the $M$-fivebrane is not wrapping around the $11^{th}$
dimension.
This is always true no matter how the sixbrane approaches to or goes
away from the branch $y_+$.
This situation means the following.
Even when a fourbrane is created in the IIA limit, this fourbrane
cannot be identified with a Dirichlet fourbrane, because it does not
create vortex at the endpoint on the NS fivebrane.

To save the situation try embedding the curve (\ref{SWc}) in a
slightly different way.
Flip the direction of the Dirac string, and carry out the analysis
from the beginning.
We should notice that changing thing is only the direction of the
Dirac string and others such as all the fourbranes and sixbranes stay
in their positions.
The positions of the two branches $y_\pm$ are exchanged, however
the radial parts of the {\it left and right} fivebrane embedding
formulae do not change.
Now, the left part of the $M$-fivebrane changes from the branch $y_+$
to the branch $y_-$, and the $\sigma$ can jump by $\pm4\pi$ when one
circles around $v=e_1$.
Therefore, in turn, the created $M$ fivebrane begins wrapping around
the $11^{th}$ dimension.
Then, in the IIA limit the endpoint of the created fourbrane on the NS 
fivebrane behaves as a vortex, and then the fourbrane may be
identified with a Dirichlet fourbrane.
One important fact is that the branch $y_-\sim v-e_1$ is always
penetrated by the Dirac string and the $\sigma$ of this branch, not
of the branch $y_+$, can jump by $\pm4\pi$.

We learn that the branch $y_-$ (or $y_+$) is not always to the right
(or left) of the $j=1^{th}$ sixbrane, depending on the direction of
the Dirac string.
When the direction of the Dirac string changes, the positions of the
two branches are also exchanged.
Unchanging things are the projected curve on the base space of the
Taub-NUT space.

\subsection*{Dirac string as a shadow of sixbrane}
Let us place the $j=1^{th}$ sixbrane in between the two NS fivebranes
with appropriate distances to the two fivebranes.
Even in this ``normal'' configuration, the Dirac string makes a vortex 
on the penetrated NS fivebrane.
One might try to avoid the Dirac string by separately considering
about each NS fivebrane, flipping the Dirac string appropriately.
However, this is impossible.
We indeed need the Dirac string, since the sixbrane contributes to the 
one-loop $\beta$-function of the low energy field theory through its
Dirac string.
We will explain this fact in the following.

First, let us briefly recall the basic ingredients needed to explain.
In general, the difference of the variables $s$ on the two side of the
$M$-fivebrane gives the one-loop gauge coupling constant
$\tau\equiv\theta/\pi + 8\pi i/g^2$, or more precisely
$i\pi\tau = (s_1-s_2)/2 = \beta_1\ln v+\mbox{constant}$ 
at large distance $v\sim\infty$.\cite{Witten}
Here, $s_1$ and $s_2$ are the positions of the left and the right
fivebranes in $s$ plane, respectively.
When one circles around the endpoint of a Dirichlet fourbrane,
$s_1$ or $s_2$ can jump by $\pm4\pi$, which counts a contribution to
the $\beta$-function.
An important fact is that this contribution persists owing to the
holomorphy for any scale of $v$.
We should notice that that the contributions are considered of the
bi-fundamental or fundamental matters from Dirichlet fourbranes or
semi-infinite Dirichlet fourbranes, but the argument is missing for
matters from Dirichlet sixbranes.

Now, we come back to the explanation.
No matter which direction is chosen for the Dirac string, one of 
the NS fivebranes is penetrated by the Dirac string, and on this side
it makes a vortex at $v=e_1$.
On the other side we have no vortex.
Owing to the Dirac string, the vortex of the penetrated NS fivebrane
contributes to the $\beta$-function coefficient $\beta_1$ by one;
$\beta_1\to\beta_1+1$, and the other NS fivebrane does not
contribute.
This is easily checked from their angular part of the embedding
formulae $y=y_\pm$.
As a result, we have the correct one-loop $\beta$-function
coefficient, for example, $\beta_1 = -2n_c+n_f$ for $SU(n_c)$ SQCD
with $n_f$ flavors, when the fundamental matters are supplied by the
sixbranes.

This result can be generalized to more complicated cases, where we
have many NS fivebranes.
Let us consider one of the many gauge couplings and calculate the
difference $s_{\alpha-1}-s_\alpha$ between the $\alpha-1^{th}$ and
the $\alpha^{th}$ NS fivebranes.
The contribution from the sixbranes in between these two NS fivebranes 
are exactly the same as that in the above $SU(n_c)$ SQCD case.
Suppose that both of these two fivebranes are not penetrated by a
Dirac string.
We have no contribution to the $\beta$-function.
Suppose that both of these two fivebranes are penetrated by a Dirac
string.
Each vortex of the two fivebranes contributes to the variable $s$,
however they cancel in the $\beta$-function.

\section*{Acknowledgments}

We would like to thank T. Nakatsu, K. Ohta and T. Yokono
for discussions and comments.
The author is supported in part by the JSPS
Research Fellowships.


\begin{thebibliography}{99}


\bibitem{HW}
A. Hanany and E. Witten,
``Type IIB Superstrings, BPS Monopoles and Three-Dimensional
Gauge Dynamics'',
{\it Nucl. Phys.} {\bf B492} (1997) 152, hep-th/9611230.

\bibitem{Anomaly}
C.P. Bachas, M.R. Douglas and M.B. Green,
``Anomalous Creation of Branes'', hep-th/9705074. \\
S.P. de Alwis,
``A Note on Brane Creation'', hep-th/9706142.

\bibitem{D0D8}
U. Danielsson, G. Ferretti and I.R. Klebanov,
``Creation of Fundamental Strings by Crossing D-branes'',
{\it Phys. Rev. Lett.} {\bf 79} (1997) 1984, hep-th/9705084. \\
O. Bergman, M.R. Gaberdiel and G. Lifschytz,
``Branes, Orientifolds and the Creation of Elementary Strings'',
hep-th/9705130.\\
P. Ho and Y. Wu,
``Brane Creation in M(atrix) theory'', hep-th/9708137.

\bibitem{D0O6} 
Y. Imamura, 
``D-particle creation on an orientifold plane'', 
hep-th/9710026. 

\bibitem{D4D4}
N. Ohta, T. Shimizu and J-G. Zhou,
``Creation of Fundamental String in M(atrix) theory'',
hep-th/9710218.

\bibitem{NOYY2}
T. Nakatsu, K. Ohta, T. Yokono and Y. Yoshida,
``A Proof of Brane Creation via $M$-theory'',
hep-th/9707258.

\bibitem{Witten}
E. Witten,
``Solutions of Four-Dimensional Field Theories via $M$ Theory'',
hep-th/9703166.

\bibitem{Townsend}
P.K. Townsend, ``The eleven-dimensional supermembrane revisited'',
{\it Phys. Lett.} {\bf B350} (1995) 184,
hep-th/9501068.

\bibitem{NOYY}
T. Nakatsu, K. Ohta, T. Yokono and Y. Yoshida,
``Higgs Branch of $N=2$ SQCD and $M$ theory Branes'',
hep-th/9707258.

\end{thebibliography}
\end{document}